\documentclass[amsmath,amssymb,aps,pra,reprint,groupedaddress,showpacs]{revtex4-1}

\usepackage{verbatim}

\newcommand{\Proof}{\noindent\textbf{Proof.}\quad}
\newcommand{\qed}{\hfill$\Box$}

\newtheorem{theorem}{Theorem}

\newtheorem{corollary}[theorem]{Corollary}

\begin{document}

\title{Block synchronization for quantum information}

\author{Yuichiro Fujiwara}
\email[]{yuichiro.fujiwara@caltech.edu}
\affiliation{Division of Physics, Mathematics and Astronomy, California Institute of Technology, MC 253-37,
Pasadena, California 91125, USA}

\date{\today}

\begin{abstract}
Locating the boundaries of consecutive blocks of quantum information
is a fundamental building block for advanced quantum computation and quantum communication systems.
We develop a coding theoretic method for properly locating boundaries of quantum information without relying on external synchronization
when block synchronization is lost.
The method also protects qubits from decoherence
in a manner similar to conventional quantum error-correcting codes,
seamlessly achieving synchronization recovery and error correction.
A family of quantum codes that are simultaneously synchronizable and error-correcting is given through this approach.
\end{abstract}

\pacs{03.67.Pp, 03.67.Hk, 03.67.Lx}

\maketitle

\section{Introduction}
The field of quantum information theory has experienced rapid and remarkable progress
toward understanding and realizing large-scale quantum computation and quantum communication.
One of the most important missions is to develop theoretical foundations for robust and reliable quantum information processing.
The discovery of the fact that it is even possible for us to correct the effects of decoherence on quantum states was
one of the most important landmarks in quantum information theory in this regard \cite{Shor}.
The field has since made various kinds of remarkable progress, from developing quantum analogues of important concepts in classical information theory
to finding surprising phenomena that are uniquely quantum information theoretic \cite{MikeandIke}.
Quantum error correction has been realized in various experiments as well \cite{CPMKLZHS,KLMN,Cetal,BVFC,Schindleretal,MBRL,RDNSFGS,TopoQEC,ZLS}.

One of the most important problems on reliable quantum information processing that remain unaddressed, however, is block synchronization
(or, more commonly, ``frame synchronization'' in the language of classical communications
\footnote{To avoid confusion with ``shared reference frames'' treated in \cite{BRS}, we use the term ``block'' as a synonym for ``frame'' throughout this paper}).
In classical digital computation and communications, virtually all data have some kind of block structure,
which means that in order for one to make sense of data, one must know the exact positions of the boundaries of each block of information, or word, in a stream of bits.

This fact will stay the same in the quantum domain.
In fact, not only will the actual quantum information one wishes to process most likely have a block structure for the same reason as in the classical domain,
but procedures for manipulating quantum information also typically demand very precise alignment.
For instance, we have a means to encode one qubit of information into five physical qubits
to reduce the effects of decoherence to the theoretical limit \cite{LMPZ}.
However, this does not mean that we can apply the procedure to, say, the last three qubits from an encoded quantum state
and the first two qubits from the following information block to correct errors. If that worked,
one would still not be able to correctly interpret the information carried by the qubits;
after all, ``quantum information theory'' is not quite the same as ``antumin formationth eory'' with ``qu'' before it.

Block synchronization is critical when correct block alignment can not be provided or is difficult to provide by a simple external mechanism.
For instance, block synchronization is a critical problem in virtually any area of classical digital communications,
where two parties are physically distant, so that synchronization must be achieved through some special signaling procedure,
such as inserting ``marker'' bits or using a specially allocated bit pattern as ``preamble'' to signal the start of each block
(see, for example, \cite{sklar,bregni} for the basics of block synchronization techniques for digital communications).

It is true that if we assume that a qubit always goes through wires as expected in a quantum circuit
and that storing, retrieval, and transmission of quantum information are always securely synchronized by external physical mechanisms,
then block synchronization is certainly not a problem.
However, such a strict assumption imposes demanding requirements on hardware and limits what quantum information processing can offer.
For example, without a software solution to block synchronization,
quantum communication would have to always be supported by perfectly synchronized classical communications to a large degree
\footnote{One may argue that because of the no-cloning theorem, quantum communication would largely be restricted to point-to-point systems
in which block synchronization is easier to establish than in broadcast networks. However, the very theorem forbids arguably the simplest and popular solution
to any kind of error including synchronization errors, that is, requesting a perfect copy of lost information.
Moreover, in quantum computing scenarios, because large-scale computing systems would benefit from internal communications between components
and/or external communications, a fundamental building block of communications would become increasingly more relevant
as the scale of realizable quantum computation grows.}.

One of the most substantial barriers to establishing block synchronization in the quantum domain is the fact that
measuring qubits usually destroys the quantum information they contain.
Existing classical block synchronization techniques typically require that
the information receiver or processing device constantly monitor the data to pick up on inserted boundary signals,
which translates into constant measurement of all qubits in the quantum case.
Hence, if an analogue of a classical synchronization scheme such as inserting preamble were to be employed in a naive manner,
one would have to know exactly where those inserted boundary signals are in order not to disturb quantum information contained in data blocks,
which would require accurate synchronization to begin with.

One might then expect that a sophisticated block synchronization scheme based on information theory would be more attractive
and promising in the quantum world.
Another big hurdle lies exactly here;
sophisticated coding for synchronization is already a notoriously difficult problem in classical information theory
(see, however, \cite{MBT} for a recent survey of coding theoretical approaches to fighting various kinds of synchronization error for the classical case).
Making things more challenging, quantum bits are thought to be more vulnerable to environmental noise than classical bits,
which implies that we ought to simultaneously answer the need for strong protection from the effects of decoherence.

The primary purpose of the present paper is to show that it is, indeed, possible to encode information about the boundaries of blocks into qubits
in such a way that block synchronization recovery and quantum error correction are seamlessly integrated.
The proposed scheme does not rely on external synchronization mechanisms or destroy quantum information by searching for boundaries.
We make use of classical error-correcting codes with certain algebraic properties,
so that the problem of finding such quantum synchronizable error-correcting codes is reduced to that of searching for special classical codes.

In the next section, we give a simple mathematical model of block synchronization in the quantum domain
and define quantum synchronizable error-correcting codes.
The details of our scheme is presented in Section \ref{synchronizable}.
Concluding remarks are given in Section \ref{sec:conclusion}.

\section{block synchronization}\label{formal}
Here we give a simple mathematical model of block synchronization in the quantum setting.
Note that while the term block might seem to suggest that each block is encoded by the same block code,
we may treat them as more general structures, so that different blocks can contain different numbers of qubits encoded by different coding schemes. 

Let $Q = (q_0,\dots, q_{x-1})$ be an ordered set of length $x$, where each element represents a qubit.
A \textit{block} $F_i$ is a set of consecutive elements of $Q$.
Let $\mathcal{F} = \{F_0, \dots, F_{y-1}\}$ be a set of blocks.
The ordered set $(Q, \mathcal{F})$ is called a \textit{block-wise structured sequence} if $|\{\bigcup_i F_i\}| = x$ and $F_i \cap F_j = \emptyset$ for $i \not= j$.
In other words, the elements of a sequence are partitioned into groups of consecutive elements called blocks.

Take a set $G = \{q_j, \dots, q_{j+g-1}\}$ of $g$ consecutive elements of $Q$.
$G$ is said to be \textit{misaligned} by $a$ qubits to the \textit{right} with respect to $(Q, \mathcal{F})$
if there exits an integer $a$ and a block $F_i$ such that $F_i = \{q_{j-a}, \dots, q_{j+g-a-1}\}$ and $G \not\in \mathcal{F}$.
If $a$ is negative, we may say that $G$ is misaligned by $\vert a \vert$ qubits to the \textit{left}.
$G$ is \textit{properly aligned} if $G \in \mathcal{F}$.

To make this mathematical model clearer, take three qubits and encode each qubit into nine qubits by Shor's nine qubit code \cite{Shor}.
The resulting 27 qubits may be seen as $Q = (q_0, \dots, q_{26})$, where the three encoded nine qubit blocks
$\left\vert \varphi_0 \right\rangle$, $\left\vert \varphi_1 \right\rangle$, and $\left\vert \varphi_2 \right\rangle$ form blocks
$F_0 = (q_0, \dots, q_8)$, $F_1 = (q_9, \dots, q_{17})$, and $F_2 = (q_{18}, \dots, q_{26})$ respectively.
These 27 qubits may be sent to a different place, stored in quantum memory or immediately processed for quantum computation.
A device, knowing the size of each information block, operates on nine qubits at a time.
If misalignment occurs by, say, two qubits to the left,
the device that tries to correct errors on qubits in $\left\vert \varphi_1 \right\rangle$ applies the error correction procedure
to the set $G$ of nine qubits $q_7$, \dots, $q_{15}$, two of which come from $F_0$ and seven of which $F_1$.
In this case, when measuring the stabilizer generator $IZZIIIIII$ of the nine qubit code to obtain the syndrome,
what the device actually does to the whole system can be expressed as
\[I^{\otimes 8}ZZI^{\otimes17}\left\vert \varphi_0 \right\rangle\left\vert \varphi_1 \right\rangle\left\vert \varphi_2 \right\rangle,\]
which, if block synchronization were correct, would be
\[I^{\otimes 10}ZZI^{\otimes15}\left\vert \varphi_0 \right\rangle\left\vert \varphi_1 \right\rangle\left\vert \varphi_2 \right\rangle.\]
$I^{\otimes 8}Z$ does not stabilize $\left\vert \varphi_0 \right\rangle$, nor does $ZI^{\otimes 8}$ $\left\vert \varphi_1 \right\rangle$.
Hence, errors are introduced to the system, rather than detected or corrected.
Similarly, if the same misalignment happens during fault-tolerant computation,
the device that tries to apply logical $\bar{X}$ to the third logical block $\left\vert \varphi_2 \right\rangle$
will apply $I^{\otimes 16}X^{\otimes 9}II$ to the 27 qubit system.

Other kinds of synchronization error such as deletion may be considered in the quantum setting
(see \cite{MBT} for mathematical models of such errors in the classical case).
As in the classical coding theory, however, we would like to separately treat them
and do not consider fundamentally different types of synchronization in the current paper.
Instead, we assume that no qubit loss or gain in the system occurs
and that a device regains access to all the qubits in proper order in the system if misalignment is correctly detected.

Our objective is to ensure that the device identifies, without destroying quantum states,
how many qubits off it is from the proper alignment should misalignment occur.
A code that is designed for detecting this type of misalignment is called a \textit{synchronizable code}
in the modern information theory literature.
Borrowing this term, we call a coding scheme
a \textit{quantum synchronizable} $(a_l, a_r)$-$[[n,k]]$ \textit{code}
if it encodes $k$ logical qubits into $n$ physical qubits and
corrects misalignment by up to $a_l$ qubits to the left and up to $a_r$ qubits to the right.

We assume that a linear combination of $I$, $X$, $Z$, and $Y$ acts on each qubit independently over a noisy quantum channel.
For error correction against such errors, we employ a version of syndrome decoding and show how to correct errors.
In principle, the true values of the minimum distances of our quantum synchronizable codes can be computed.
However, we focus on how many nontrivial quantum errors our decoding procedure can correct.
Hence, the actual minimum distances of our quantum synchronizable codes may be larger than
what our decoding algorithm suggests.

In what follows, we give a general construction for quantum synchronizable error-correcting codes
and describe the procedures of encoding, error correction, synchronization recovery, and decoding.
An infinite class of such quantum codes will be given at the end of the next section as an example.

\section{Coding scheme}\label{synchronizable}
In this section we give the mathematical details of our solution and show how to realize quantum synchronizable codes.
We employ classical and quantum coding theory.
For the basic facts and notions in classical and quantum coding theories, the reader is referred to \cite{HP,MikeandIke}.

\subsection{Preliminaries}\label{subsec:pre}
As usual, we define a binary linear $[n,k]$ code as a $k$-dimensional subspace of $\mathbb{F}^n_2$,
the $n$-dimensional vector space over the binary field.
Because we do not consider a code over another field, we always assume that a classical code is binary unless otherwise stated.

A \textit{cyclic} code $\mathcal{C}$ is a linear $[n,k]$ code with the property that
if $\boldsymbol{c} = (c_0,\dots, c_{n-1})$ is a codeword of $\mathcal{C}$, then so is every cyclic shift of $\boldsymbol{c}$.
It is known that, by regarding each codeword as the coefficient vector of a polynomial in $\mathbb{F}_2[x]$,
a cyclic code can be seen as a principal ideal in the ring $\mathbb{F}_2[x]/(x^n-1)$
generated by the unique monic nonzero polynomial $g(x)$ of minimum degree in the code which divides $x^n-1$.
Computations in $\mathbb{F}_2[x]/(x^n-1)$ are modulo $x^n-1$.
A cyclic shift thus corresponds to multiplying by $x$, and the code can be written as
$\mathcal{C} = \{i(x)g(x) \ \vert \ \deg(i(x)) < k\}$.
Multiplying by $x$ is an automorphism.
The orbit of a given codeword $i(x)g(x)$ by this group action is written as $Orb(i(x)g(x)) = \{i(x)g(x), xi(x)g(x), x^2i(x)g(x), \dots\}$.

Let $\mathcal{C}$ and $\mathcal{D}$ be two linear codes of the same length.
$\mathcal{D}$ is $\mathcal{C}$-\textit{containing} if $\mathcal{C} \subseteq \mathcal{D}$.
It is \textit{dual-containing} if it contains its dual $\mathcal{D}^{\perp} = \{\boldsymbol{d}^{\perp} \in \mathbb{F}_2^n \ \vert\ 
\boldsymbol{d} \cdot \boldsymbol{d}^{\perp} = \boldsymbol{0}, \boldsymbol{d} \in \mathcal{D}\}$.
The Calderbank-Shor-Steane construction \cite{CS,Steane} turns a $\mathcal{C}$-containing linear code
into a quantum error-correcting code, called a \textit{CSS} code.
If we apply a dual-containing $[n,k,d]$ linear code,
the resulting CSS code is of parameters $[[n, 2k-n,d']]$ for some $d' \geq d$.
In terms of block synchronization, this CSS code is a quantum synchronizable $(0,0)$-$[[n,2k-n]]$ code,
as the code tolerates no synchronization error.
Any combination of up to $\lfloor \frac{d-1}{2} \rfloor$ quantum errors can be corrected through a separate two-step error correction procedure
by directly exploiting the error correction mechanism of the corresponding classical code.
A higher quantum error correction capability may be achieved if the code is degenerate.
For the sake of simplicity, however, we do not investigate the degeneracy of each individual quantum error-correcting code.
In the remainder of this paper, we assume familiarity with the structure of CSS codes as well as their basic encoding and decoding mechanisms
given in a standard textbook such as \cite{MikeandIke}.

\subsection{Main theorem}\label{subsec:mainTh}
Our main theorem employs a pair of cyclic codes $\mathcal{C}$ and $\mathcal{D}$ satisfying $\mathcal{C}^{\perp} \subseteq \mathcal{C} \subset \mathcal{D}$
to generate a quantum synchronizable code.

\begin{theorem}\label{main}
If there exist a dual-containing cyclic $[n, k_1,d_1]$ code $\mathcal{C}$ and a $\mathcal{C}$-containing cyclic $[n, k_2, d_2]$ code with $k_1 < k_2$,
then for any pair of nonnegative integers $a_l$, $a_r$ satisfying $a_l + a_r < k_2 - k_1$
there exists a quantum synchronizable $(a_l, a_r)$-$[[n+a_l+a_r, 2k_1-n]]$ code that corrects
at least up to $\lfloor\frac{d_1-1}{2}\rfloor$ phase errors and
at least up to $\lfloor\frac{d_2-1}{2}\rfloor$ bit errors.
\end{theorem}

To prove Theorem \ref{main}, we realize a quantum synchronizable code
as a carefully translated vector space similar to a CSS code.
The proof of the above theorem will be completed in Section \ref{PTh1} after describing encoding and decoding procedures
in Section \ref{enc} and Sections \ref{1}--4.

Let $\mathcal{C}$ be a dual-containing cyclic $[n, k_1,d_1]$ code that lies in another cyclic $[n, k_2, d_2]$ code $\mathcal{D}$ with $k_1 < k_2$.
Define $g(x)$ as the the generator of $\mathcal{D} = \left\langle g(x) \right\rangle$
which is the unique monic nonzero polynomial of minimum degree in $\mathcal{D}$.
Define also $h(x)$ as the generator of $\mathcal{C}$ which is the unique monic nonzero polynomial of minimum degree in $\mathcal{C}$.
Since $\mathcal{C} \subset \mathcal{D}$,
the generator $g(x)$ divides every codeword of $\mathcal{C}$.
Hence, $h(x)$ can be written as $h(x) = f(x)g(x)$ for some polynomial $f(x)$ of degree $n-k_1-\deg(g(x)) = k_2-k_1$.

For every polynomial $j(x) = j_0 + j_1x +\cdots+j_{n-1}x^{n-1}$ of degree less than $n$,
define $\left\vert j(x) \right\rangle$ as the $n$ qubit quantum state
$\left\vert j(x) \right\rangle = \left\vert j_0 \right\rangle\left\vert j_1 \right\rangle\cdots\left\vert j_{n-1} \right\rangle$.
For a set $J$ of polynomials of degree less than $n$, we define $\left\vert J \right\rangle$ as
\[\left\vert J \right\rangle = \frac{1}{\left\vert J \right\vert}\sum_{j(x) \in J}\left\vert j(x) \right\rangle.\]
For a polynomial $k(x)$, we define $J + k(x) = \{j(x)+k(x) \ \vert \ j(x) \in J\}$.

Let $R = \{r_i(x)\ \vert \ 0 \leq i \leq 2^{2k_1-n-1}\}$ be a system of representatives of the cosets $\mathcal{C}/\mathcal{C}^{\perp}$.
Consider the set $V_g = \left\{\left\vert \mathcal{C}^{\perp} + r_i(x) + g(x) \right\rangle \ \vert \ r_i(x) \in R\right\}$ of $2^{2k_1-n}$ states.
Because $R$ is a system of representatives,
these $2^{2k_1-n}$ states form an orthonormal basis.
Let $\mathcal{V}_g$ be the vector space of dimension $2^{2k_1-n}$ spanned by $V_g$.
We employ this translated space $\mathcal{V}_g$ to prove Theorem \ref{main}.

\subsection{Encoding}\label{enc}
Take a full-rank parity-check matrix $H_{\mathcal{D}}$ of $\mathcal{D}$.
For each row of $H_{\mathcal{D}}$, replace zeros with $I$s and ones with $X$s.
Perform the same replacement with $I$s for zeros and $Z$s for ones.
Because $\mathcal{C}^{\perp}\subset\mathcal{C}\subset\mathcal{D}$ implies $\mathcal{D}^{\perp}\subset\mathcal{D}$,
the code $\mathcal{D}$ is a dual-containing cyclic code of dimension $k_2$.
Hence, the resulting $2(n-k_2)$ Pauli operators on $n$ qubits form stabilizer generators $\mathcal{S}_{\mathcal{D}}$ of the Pauli group on $n$ qubits
that fixes a subspace of dimension $2^{k_2}$.
The set of the Pauli operators on $n$ qubits in $\mathcal{S}_{\mathcal{D}}$ that consist of only $Z$s and $I$s is referred to as $\mathcal{S}_{\mathcal{D}}^Z$.
Construct stabilizer generators $\mathcal{S}_{\mathcal{C}}$ in the same manner by using $\mathcal{C}$.

Take an arbitrary $2k_1-n$ qubit state $\left\vert \varphi \right\rangle$, which is to be encoded.
By using an encoder for the CSS code of parameters $[[n,2k_1-n]]$ defined by $\mathcal{S}_{\mathcal{C}}$,
the state $\left\vert \varphi \right\rangle$ is encoded into $n$ qubit state
$\left\vert \varphi \right\rangle_{\text{enc}} = \sum_i\alpha_i\left\vert \boldsymbol{v}_i \right\rangle$,
where each $\boldsymbol{v}_i$ is an $n$-dimensional vector
with the orthogonal basis being $\left\{\left\vert \mathcal{C}^{\perp} + r_i(x) \right\rangle \ \vert \ r_i(x) \in R\right\}$.
Let $U_{\boldsymbol{g}}$ be the unitary operator that adds the coefficient vector $\boldsymbol{g}$ of $g(x)$.
By applying $U_{\boldsymbol{g}}$, we have:
\[U_{\boldsymbol{g}} \left\vert \varphi \right\rangle_{\text{enc}} = \sum_i\alpha_i\left\vert \boldsymbol{v}_i +\boldsymbol{g}, \right\rangle.\]

Take a pair of nonnegative integers $a_l$, $a_r$ that satisfy $a_l+a_r < k_2-k_1$.
Using $a_l+a_r$ ancilla qubits and CNOT gates, we take this state to an $n+a_l+a_r$ qubit state as follows:
\[\left\vert 0\right\rangle^{\otimes a_l}U_{\boldsymbol{g}}\left\vert \varphi \right\rangle_{\text{enc}}\left\vert 0\right\rangle^{\otimes a_r}
\rightarrow \sum_i\alpha_i\left\vert \boldsymbol{w}_i^1, \boldsymbol{v}_i+\boldsymbol{g}, \boldsymbol{w}_i^2 \right\rangle,\]
where $\boldsymbol{w}_i^1$ and $\boldsymbol{w}_i^2$ are the last $a_l$
and the first $a_r$ portions of the vector $\boldsymbol{v}_i +\boldsymbol{g}$ respectively.
The resulting state
$\left\vert \psi \right\rangle_{\text{enc}} = \sum_i\alpha_i\left\vert \boldsymbol{w}_i^1, \boldsymbol{v}_i+\boldsymbol{g}, \boldsymbol{w}_i^2 \right\rangle$
then goes through a noisy quantum channel.

\subsection{Error correction and block synchronization}\label{D}
Gather $n+a_l+a_r$ consecutive qubits $G = (q_0,\dots, q_{n+a_l+a_r-1} )$.
We assume the situation where correct block synchronization means that $G$ is exactly the qubits of $\left\vert \psi \right\rangle_{\text{enc}}$,
but $G$ can be misaligned by $a$ qubits to the right, where $-a_l \leq a \leq a_r$.

Let $P = (p_0,\dots, p_{n+a_l+a_r-1})$ be the $n+a_l+a_r$ qubits of the encoded state.
If $a=0$, then $P = G$.
Define $G_m = (q_{a_l}, \dots, q_{a_l + n -1})$.
By assumption, $G_m = (p_{a_l + a}, \dots, p_{a_l + n -1 + a})$.
Let $n$-fold tensor product $E$ of linear combinations of the Pauli matrices
be the errors that occurred on $P$.

We first outline the bit error correction procedure on the window $G_m$.
Synchronization is recovered after making $G_m$ free from bit errors.
The bit errors outside of $G_m$ are then corrected.
The phase errors on qubits will be treated at the final step after reversing the extension process.

\subsubsection{Bit error correction on the initial window}\label{1}
We correct bit errors that occurred on qubits in $G_m$ in the same manner as the separate two-step error correction procedure for a CSS code.
Since $\mathcal{C} \subset \mathcal{D}$,
the vector space spanned by the orthogonal basis stabilized by $\mathcal{S}_{\mathcal{D}}$ contains $\mathcal{V}_g$ as a subspace.
Hence, through a unitary transformation using $\mathcal{S}_{\mathcal{D}}^Z$, we can obtain the error syndrome in the same manner
as when detecting errors with the CSS code defined by $\mathcal{S}_{\mathcal{D}}$ as follows:
\[E\left\vert \psi \right\rangle_{\text{enc}}\left\vert 0 \right\rangle^{\otimes n-k_2}\\
\rightarrow E\left\vert \psi \right\rangle_{\text{enc}}\left\vert \chi \right\rangle,\]
where $\left\vert \chi \right\rangle$ is the $n-k_2$ qubit syndrome by $\mathcal{S}_{\mathcal{D}}^Z$.
If $E$ introduced at most $\lfloor \frac{d_2-1}{2} \rfloor$ bit errors on qubits in $G_m$,
these quantum errors are detected and then corrected by applying the $X$ operators if necessary.

More formally, rewrite the original encoded state
$\left\vert \psi \right\rangle_{\text{enc}} = \sum_i\alpha_i\left\vert \boldsymbol{w}_i^1, \boldsymbol{v}_i+\boldsymbol{g}, \boldsymbol{w}_i^2 \right\rangle$
as
\[\left\vert \psi \right\rangle_{\text{enc}} = \sum_i\alpha_i\left\vert \boldsymbol{l}_i, \boldsymbol{c}_i, \boldsymbol{r}_i \right\rangle,\]
where $\boldsymbol{c}_i$ correspond to the window misaligned by $a$ qubits to the right,
which can be obtained by cyclically shifting $\boldsymbol{v}_i+\boldsymbol{g}$.
Hence, the binary vectors $\boldsymbol{l}_i,$ and $\boldsymbol{r}_i$ are of lengths $a_l+a$ and $a_r-a$ respectively.

Without loss of generality, we consider $E$
the discretized bit errors and phase errors on the $n+a_l+a_r$ qubits of $\left\vert \psi \right\rangle_{\text{enc}}$.
Let $\boldsymbol{e}^b$ be the $(n+a_l+a_r)$-dimensional binary error vector such that
$i \in \text{supp}(\boldsymbol{e}^b)$ if and only if a bit error occurred on qubit $p_i$.
In other words, the positions of $1$s in $\boldsymbol{e}^b$ represent which qubits are bitwise flipped.
Define the phase error vector $\boldsymbol{e}^p$ in the same way for the phase errors that occurred on $\left\vert \psi \right\rangle_{\text{enc}}$.
Then, the transformation due to the noisy quantum channel that introduced quantum error $E$ is
\begin{align*}
\left\vert \psi \right\rangle_{\text{enc}} &\rightarrow E\left\vert \psi \right\rangle_{\text{enc}}\\
&=  \sum_i\alpha_i(-1)^{(\boldsymbol{l}_i, \boldsymbol{c}_i, \boldsymbol{r}_i)\cdot \boldsymbol{e}^p}
\left\vert (\boldsymbol{l}_i, \boldsymbol{c}_i, \boldsymbol{r}_i)+\boldsymbol{e}^b \right\rangle.
\end{align*}

Write the bit error vector as $\boldsymbol{e}^b = (\boldsymbol{e}^b_l,\boldsymbol{e}^b_c,\boldsymbol{e}^b_r)$,
where $\boldsymbol{e}^b_l$, $\boldsymbol{e}^b_c$, and $\boldsymbol{e}^b_r$ are
the first $a_l+a$, next $n$, and last $a_r-a$ bits of $\boldsymbol{e}^b$ respectively.
Recall that $H_{\mathcal{D}}$ is the full-rank parity-check matrix of $\mathcal{D}$ corresponding to the stabilizer generators.
We perform the following unitary transformation using $\mathcal{S}_{\mathcal{D}}^Z$ with $n-k_2$ ancilla qubits:
\[E\left\vert \psi \right\rangle_{\text{enc}}\left\vert 0 \right\rangle^{\otimes n-k_2}\\
\rightarrow E\left\vert \psi \right\rangle_{\text{enc}}\left\vert H_{\mathcal{D}}\boldsymbol{e}^b_c \right\rangle.\]
Because $H_{\mathcal{D}}$ is a parity-check matrix of $\mathcal{D}$,
measuring the ancilla gives the error syndrome in the same manner as the corresponding classical linear code does.
Thus, as in the standard bit error correction procedure for a CSS code,
if we assume that $E$ introduced at most $\lfloor \frac{d_2-1}{2} \rfloor$ bit errors on qubits in $G_m$,
applying $X$ operators to qubits specified by the error syndrome $H_{\mathcal{D}}\boldsymbol{e}^b_c$ takes the encoded sate with errors to
\[E'\hspace{-0.05mm}\left\vert \psi \right\rangle_{\text{enc}}\hspace{-0.15mm}=\hspace{-0.15mm}
\sum_i\hspace{-0.15mm}\alpha_i(-1)^{(\boldsymbol{l}_i, \boldsymbol{c}_i, \boldsymbol{r}_i)\cdot \boldsymbol{e}^p}\hspace{-0.2mm}
\left\vert (\boldsymbol{l}_i, \boldsymbol{c}_i, \boldsymbol{r}_i)\hspace{-0.15mm}+\hspace{-0.15mm}
(\boldsymbol{e}^b_l,\boldsymbol{0},\boldsymbol{e}^b_r) \right\rangle\hspace{-0.4mm},\]
where $E'$ represents the partially corrected quantum errors.

\subsubsection{Synchronization recovery}\label{2}
We perform synchronization recovery by exploiting the bit-error-free $G_m$ we just obtained.
Recall that all codewords of $\mathcal{C}^{\perp}$ and $r_i(x) \in R$ belong to $\mathcal{C}$, and hence to $\mathcal{D}$ as well.
Because $g(x)$ is the generator of $\mathcal{D}$,
it divides any polynomial of the form $s(x) + r_i(x) + g(x)$ over $\mathbb{F}_2[x]/(x^n-1)$, where $s(x) \in \mathcal{C}^{\perp}$.
Since we have
\[s(x) + r_i(x) + g(x) = i_0(x)f(x)g(x)+i_1(x)f(x)g(x)+g(x)\]
for some polynomials $i_0(x)$ and $i_1(x)$ of degree less than $k_1$,
the quotient is of the form $j(x)f(x)+1$ for some polynomial $j(x)$.
Dividing the quotient by $f(x)$ gives $1$ as the remainder.
Note that $g(x)$ is a monic polynomial of degree $n-k_2$ that divides $x^n-1$,
where $k_2$ is strictly larger than $\lceil \frac{n}{2} \rceil$.
Let $i$ be an integer satisfying $1 \leq i \leq \lceil\frac{n}{2}\rceil \leq k_2-1$.
Then
\[\deg(x^ig(x)) = n-k_2+i \not= \deg(g(x)).\]
Hence, we have $\left\vert Orb(g(x)) \right\vert \geq k_2 > \lceil\frac{n}{2}\rceil$.
Because $\left\vert Orb(g(x)) \right\vert$ must divide $n$, we have $\left\vert Orb(g(x)) \right\vert = n$.
Thus, applying the same two-step division procedure to any polynomial appearing as a state
in cyclically shifted $V_g$ by $a$ qubits gives $x^a \pmod{f(x)}$ as the remainder.
By assumption, we have
\[0 < a_l+a_r < k_2-k_1 = \deg(f(x))\]
and $-a_l \leq a \leq a_r$.
Thus, the remainder $x^a \pmod{f(x)}$ is unique to each possible value of $a$.

Recall that every state in $V_g$ is of the form $\left\vert \mathcal{C}^{\perp} + r_i(x) + g(x) \right\rangle$.
Because $G_m$ contains no bit errors,
the basis states of the corresponding portion in $E'\left\vert \psi \right\rangle_{\text{enc}}$ are the cyclically shifted coefficient vectors of the correct polynomials.
Let $Dq_{t(x)}$ and $Dr_{t(x)}$ be the polynomial division operations on $n$ qubits that give
the quotient and remainder respectively through quantum shift registers defined by a polynomial $t(x)$ of degree less than $n$ \cite{GB}
(see also \cite{Wilde} for an alternative way to implement quantum shift registers).
Let $\mathfrak{Q} = I^{\otimes a_l + a}Dq_{g(x)}I^{\otimes a_r - a}$
and $\mathfrak{R} = I^{\otimes n + a_l + a_r}Dr_{f(x)}$,
so that the two represent applying $Dq_{g(x)}$ to the window and $Dr_{f(x)}$ to the ancilla qubits of $Dq_{g(x)}$ that contain the calculated quotient.
These operations give the syndrome for the synchronization error as
\[E'\left\vert \psi \right\rangle_{\text{enc}} \left\vert 0 \right\rangle^{\otimes n}
\xrightarrow{\mathfrak{R}\mathfrak{Q}}
E'\left\vert \psi \right\rangle_{\text{enc}} \left\vert x^a\ (\text{mod}\ {f(x)})\right\rangle,\]
where $\left\vert 0 \right\rangle^{\otimes n}$ is the ancilla for $Dq_{g(x)}$.
Hence, by regarding the remainder $x^a \pmod{f(x)}$ as the syndrome of synchronization error $a$, the magnitude and direction are identified.

\subsubsection{Bit error correction outside the initial window}\label{3}
Because we obtained the information about
how many qubits $G = (q_0,\dots, q_{n+a_l+a_r-1} )$ is away from the proper position $P = (p_0,\dots, p_{n+a_l+a_r-1})$ and in which direction,
by assumption, we can correctly shift the window to the last $n$ qubits $(p_{a_l+a_r},\dots,p_{n+a_l+a_r-1})$ of $P$.
Note that if $a$ is negative, the last $\vert a\vert$ qubits are outside of $G$,
which means that the receiver may be required to gather $\vert a\vert$ more qubits in addition to the consecutive $n+a_l+a_r$ qubits initially received.
Because we employed classical cyclic codes,
the same error correction procedure can be performed on $(p_{a_l+a_r},\dots,p_{n+a_l+a_r-1})$,
allowing for correcting bit errors that may have occurred on the last $n$ qubits of $P$.
By the same token, moving the window to the first $n$ qubits of $P$ enables us to correct the remaining bit errors on $P$.
Thus, if the channel introduced at most $\lfloor \frac{d_2-1}{2} \rfloor$ bit errors on any consecutive $n$ quibits,
we can correct all bit errors that occurred on $P$ to obtain $E'' \left\vert \psi \right\rangle_{\text{enc}}$, where $E''$ only introduces phase errors.

\subsubsection{Phase error correction}\label{4}
Next we correct the effect of the phase errors that occurred on qubits in $P$.
The first step we take is to reverse the extension operation and the unitary operation $U_{\boldsymbol{g}}$
that transformed the $n$ qubit encoded state $\left\vert \varphi \right\rangle_{\text{enc}} = \sum_i\alpha_i\left\vert \boldsymbol{v}_i \right\rangle$
into the $n+a_l+a_r$ qubit state
$\left\vert \psi \right\rangle_{\text{enc}} = \sum_i\alpha_i\left\vert \boldsymbol{w}_i^1, \boldsymbol{v}_i+\boldsymbol{g}, \boldsymbol{w}_i^2 \right\rangle$.
Here we straightforwardly apply the same CNOT operations to the qubits in $E''\left\vert \psi \right\rangle_{\text{enc}}$
as we did when extending $U_{\boldsymbol{g}}\left\vert \varphi \right\rangle_{\text{enc}}$,
then discard the $a_l+a_r$ qubits that were initially ancilla qubits for extension,
and finally apply $U_{\boldsymbol{g}}$ again to the resulting $n$ qubit state.

Write the phase error vector as $\boldsymbol{e}^p = (\boldsymbol{e}^p_{l},\boldsymbol{e}^p_{c},\boldsymbol{e}^p_{r})$,
where the binary error vectors $\boldsymbol{e}^p_{l}$, $\boldsymbol{e}^p_{c}$, and $\boldsymbol{e}^p_{r}$ correspond to the phase errors
that occurred on the first $a_l$, next $n$ and last $a_r$ qubits of $P$.
Then the above reversing operation can be described by the following transformation:
\begin{align*}
E''\left\vert \psi \right\rangle_{\text{enc}} &\rightarrow \sum_i\alpha_i
(-1)^{(\boldsymbol{v}_i+\boldsymbol{g})\cdot(\boldsymbol{e}^p_{c}+(\boldsymbol{0},\boldsymbol{e}^p_{l})+(\boldsymbol{e}^p_{r},\boldsymbol{0}))}
\left\vert \boldsymbol{v}_i \right\rangle\\
&=  e^{i\theta}\sum_i\alpha_i
(-1)^{\boldsymbol{v}_i\cdot(\boldsymbol{e}^p_{c}+(\boldsymbol{0},\boldsymbol{e}^p_{l})+(\boldsymbol{e}^p_{r},\boldsymbol{0}))}
\left\vert \boldsymbol{v}_i \right\rangle,
\end{align*}
where $\theta$ is some multiple of $\pi$,
and $(\boldsymbol{0},\boldsymbol{e}^p_{l})$ and $(\boldsymbol{e}^p_{r},\boldsymbol{0})$ are the $n$-dimensional binary vectors obtained by
padding $n-a_l$ and $n-a_r$ zeros to the head of $\boldsymbol{e}^p_{l}$ and the tail of $\boldsymbol{e}^p_{r}$ respectively.
Note that by writing as $n_p$ the number of qubits on which the phase errors occurred among the $n+a_l+a_r$ qubits,
we have
\begin{align*}
\vert  \text{supp}(\boldsymbol{e}^p_{c}&+(\boldsymbol{0},\boldsymbol{e}^p_{l})+(\boldsymbol{e}^p_{r},\boldsymbol{0}))\vert\\
&\leq \vert  \text{supp}(\boldsymbol{e}^p_{c})\vert
+ \vert \text{supp}((\boldsymbol{0},\boldsymbol{e}^p_{l}))\vert
+ \vert \text{supp}((\boldsymbol{e}^p_{r},\boldsymbol{0}))\vert\\
&= n_p.
\end{align*}
The encoded state $\left\vert \varphi \right\rangle_{\text{enc}}$ is stabilized by $\mathcal{S}_{\mathcal{C}}$.
Thus, ignoring the global phase factor $e^{i\theta}$,
if $n_p \leq \lfloor\frac{d_1-1}{2}\rfloor$, we can correctly diagnose the effect of
$\boldsymbol{e}^p_{c}+(\boldsymbol{0},\boldsymbol{e}^p_{l})+(\boldsymbol{e}^p_{r},\boldsymbol{0})$
through the standard phase error correction procedure for the CSS code based on the dual-containing cyclic code $\mathcal{C}$:
\begin{align*}
E'''\left\vert \varphi \right\rangle_{\text{enc}}&{}\left\vert 0 \right\rangle^{\otimes n-k_1}\\
&\rightarrow E'''\left\vert \varphi \right\rangle_{\text{enc}}\left\vert H_{\mathcal{C}}(\boldsymbol{e}^p_{c}+(\boldsymbol{0},\boldsymbol{e}^p_{l})+(\boldsymbol{e}^p_{r},\boldsymbol{0})) \right\rangle,
\end{align*}
where $H_{\mathcal{C}}$ is a full-rank parity-check matrix of $\mathcal{C}$
and $E'''$ is the phase error operator on $\left\vert \varphi \right\rangle_{\text{enc}}$ that represents the effect of
$\boldsymbol{e}^p_{c}+(\boldsymbol{0},\boldsymbol{e}^p_{l})+(\boldsymbol{e}^p_{r},\boldsymbol{0})$.
Applying $Z$ operators on the qubits specified by the syndrome completes the error correction procedure.

\subsubsection{Proof of Theorem \ref{main} and example codes}\label{PTh1}
We are now able to prove Theorem \ref{main}.

\noindent
\textbf{Proof of Theorem \ref{main}.}
Take a dual-containing cyclic $[n,k_1,d_1]$ code $\mathcal{C}$ that is contained in a cyclic $[n,k_2,d_2]$ code, where $k_1 < k_2$.
Encode $2k_1-n$ logical qubits into $n+a_l+a_r$ physical qubits as described above.
The error correction and synchronization recovery procedures described above correct
misalignment by $a$ qubits to the right as long as $a$ lies in the range $-a_l \leq a \leq a_r$
and correct
up to $\lfloor\frac{d_1-1}{2}\rfloor$ phase errors on the $n+a_l+a_r$ qubits and
up to $\lfloor\frac{d_2-1}{2}\rfloor$ bit errors on any consecutive $n$ qubits.
The final decoding step is completed by reducing the state $\left\vert \varphi \right\rangle_{\text{enc}} = \sum_i\alpha_i\left\vert \boldsymbol{v}_i \right\rangle$
to the original state $\left\vert \varphi \right\rangle$ by a decoding circuit of the CSS code based on the dual-containing cyclic code $\mathcal{C}$.
Thus, the scheme is a quantum synchronizable $(a_l, a_r)$-$[[n+a_l+a_r,2k_1-n]]$ code with the desired error correction capability.
\qed

To take full advantage of Theorem \ref{main}, we need dual-containing cyclic codes
that achieve large minimum distance and contain dual-containing cyclic codes of smaller dimension.
A class of the well-known Bose-Chaudhuri-Hocquenghem (BCH) codes \cite{HP} gives such classical codes.
The dual-containing properties of BCH codes have been thoroughly investigated in \cite{Steane2,AKS}.
The following is an infinite series of quantum synchronizable error-correcting codes
based on a class of such codes, called the \textit{primitive, narrow-sense} BCH codes
(see \cite{HP} for the definition and basic properties of primitive, narrow-sense BCH codes):
\begin{corollary}\label{coro}
Let $n$, $d_1$, and $d_2$ be odd integers satisfying $n=2^m -1$ and $3 \leq d_2 < d_1 \leq 2^{\lceil \frac{m}{2} \rceil}-1$,
where $m \geq 5$.
Then for some $d_1' \geq d_1$, some $d_2' \geq d_2$, and any pair of nonnegative integers $a_l$, $a_r$ satisfying $a_l+a_r < \frac{m(d_1-d_2)}{2}$
there exists a quantum synchronizable $(a_l, a_r)$-$[[n+a_l+a_r,n-m(d_2-1)]]$ code that corrects
up to $\frac{d_1'-1}{2}$ phase errors on the $n+a_l+a_r$ qubits and
up to $\frac{d_2'-1}{2}$ bit errors on any consecutive $n$ qubits.
\end{corollary}
\Proof
Let $n$, $d_1$, and $d_2$ be integers satisfying the condition given in the statement.
Let $\mathcal{D}$ be a primitive narrow-sense BCH code of length $n$
and designed distance $d_2$ such that $3 \leq d_2 < 2^{\lceil \frac{m}{2} \rceil} -1$.
Construct a primitive narrow-sense BCH code $\mathcal{C}$ by joining one or more cyclotomic cosets,
so that its designed distance $d_1$ is larger than $d_2$ but smaller than or equal to $2^{\lceil \frac{m}{2} \rceil} -1$.
The dimensions of $\mathcal{C}$ and $\mathcal{D}$ are $n-\frac{m(d_1-1)}{2}$ and $n-\frac{m(d_2-1)}{2}$ respectively.
$\mathcal{D}$ contains $\mathcal{C}$,
and the two cyclic codes are both dual-containing (see \cite{Steane2}),
forming the desired quantum synchronizable codes.
\qed

\section{Conclusion}\label{sec:conclusion}
We developed a coding scheme that seamlessly integrates block synchronization and quantum error correction.
A close relation is found between quantum synchronizable error-correcting codes and pairs of cyclic codes with special properties.
Through this relation, the well-known BCH codes were shown to generate desirable quantum codes for block synchronization.

In classical communications, a unified method for synchronization and error correction
can reduce implementation complexity \cite{CSA}.
A similar method using cyclic codes has also been proposed recently in the classical domain
for simple implementation of asynchronous code division multiple access (CDMA) systems with random delays \cite{WC}.
We hope that our seamlessly unified solution to block synchronization and quantum error correction
may help simplify requirements on hardware and
open up new possibilities of quantum computation and quantum communication
such as transmission of a large amount of consecutive quantum information blocks with little aid from classical communications.

One potential weakness of the approach presented in this paper is that
our quantum synchronizable codes of length $n+a_l+a_r$ may face a larger number of quantum errors
than the underlying standard CSS codes of length $n$ would because of their extended lengths.
For instance, in a scenario where the receiver missed the first several qubits,
the window may be suffering from severe quantum errors which may not be correctable.
Phase error correction requires particular attention in this regard because while
the current scheme takes advantage of the subcode $\mathcal{C}$, which typically has a larger minimum distance than $\mathcal{D}$ for bit errors,
the error correction scheme for phase errors is expected to handle all phase errors at once unlike the bit error correction procedure.
While the ability to recover from misalignment is highly valuable because even the slightest synchronization error is fatal to information transmission,
these weaknesses should be noted and are worthy of further investigation.

One aspect we may be able to improve is the maximum magnitude of a correctable synchronization error.
The scheme presented in this paper relies on the uniqueness of the syndrome for each possible combination of the magnitude and direction.
While the remainder $x^a \pmod{f(x)}$ after the two-step division procedure for synchronization recovery is certainly unique
if we limit $a_l+a_r$ to be less than $\deg(f(x))$, this may be overly conservative in a sense.
In fact, there are $2^{\deg(f(x))}$ possible polynomials of degree $\deg(f(x))$ or smaller
while we only need at most $n$ distinct synchronization error syndromes
even if we extend a CSS code of length $n$ to a full $2n$ qubit code by copying all $n$ qubits with CNOT gates.
While our scheme does not appear to allow a better general bound on the maximum correctable magnitude
in a simple form without a deeper observation and careful modification,
it is plausible that a sophisticate treatment of syndromes
may yield quantum synchronizable codes with better synchronization error tolerance than is proved in this paper.

Finally, while we have focused on binary dual-containing cyclic codes,
it is certainly of interest to look into more general approaches to quantum error correction such as orthogonal pairs of cyclic codes
that are not dual-containing and the quantum error-correcting codes from additive codes over $\mathbb{F}_4$ found in \cite{CRSSIEEE}.
While CSS codes and similar quantum error-correcting codes based on classical cyclic codes
that admit decoding through quantum shift registers have not been studied very well in the literature,
there are some examples that have very similar structures such as quantum Reed-Solomon codes \cite{GGB}
(see also \cite{GB} for a possible decoding scheme for this type of quantum error-correcting code through quantum shift registers).
A further look into these types of quantum cyclic code would be of interest.

\begin{acknowledgments}
YF thanks V.D.\ Tonchev, D. Clark, M.M.\ Wilde, and R.M.\ Wilson for constructive comments and stimulating discussions.
He is grateful to the anonymous referees for their careful reading and valuable comments.
A significant portion of Section \ref{D} is due to their suggestions.
This work is supported by JSPS.
\end{acknowledgments}


\begin{thebibliography}{29}%
\makeatletter
\providecommand \@ifxundefined [1]{%
 \@ifx{#1\undefined}
}%
\providecommand \@ifnum [1]{%
 \ifnum #1\expandafter \@firstoftwo
 \else \expandafter \@secondoftwo
 \fi
}%
\providecommand \@ifx [1]{%
 \ifx #1\expandafter \@firstoftwo
 \else \expandafter \@secondoftwo
 \fi
}%
\providecommand \natexlab [1]{#1}%
\providecommand \enquote  [1]{``#1''}%
\providecommand \bibnamefont  [1]{#1}%
\providecommand \bibfnamefont [1]{#1}%
\providecommand \citenamefont [1]{#1}%
\providecommand \href@noop [0]{\@secondoftwo}%
\providecommand \href [0]{\begingroup \@sanitize@url \@href}%
\providecommand \@href[1]{\@@startlink{#1}\@@href}%
\providecommand \@@href[1]{\endgroup#1\@@endlink}%
\providecommand \@sanitize@url [0]{\catcode `\\12\catcode `\$12\catcode
  `\&12\catcode `\#12\catcode `\^12\catcode `\_12\catcode `\%12\relax}%
\providecommand \@@startlink[1]{}%
\providecommand \@@endlink[0]{}%
\providecommand \url  [0]{\begingroup\@sanitize@url \@url }%
\providecommand \@url [1]{\endgroup\@href {#1}{\urlprefix }}%
\providecommand \urlprefix  [0]{URL }%
\providecommand \Eprint [0]{\href }%
\providecommand \doibase [0]{http://dx.doi.org/}%
\providecommand \selectlanguage [0]{\@gobble}%
\providecommand \bibinfo  [0]{\@secondoftwo}%
\providecommand \bibfield  [0]{\@secondoftwo}%
\providecommand \translation [1]{[#1]}%
\providecommand \BibitemOpen [0]{}%
\providecommand \bibitemStop [0]{}%
\providecommand \bibitemNoStop [0]{.\EOS\space}%
\providecommand \EOS [0]{\spacefactor3000\relax}%
\providecommand \BibitemShut  [1]{\csname bibitem#1\endcsname}%
\let\auto@bib@innerbib\@empty
\bibitem [{\citenamefont {Shor}(1995)}]{Shor}%
  \BibitemOpen
  \bibfield  {author} {\bibinfo {author} {\bibfnamefont {P.~W.}\ \bibnamefont
  {Shor}},\ }\href@noop {} {\bibfield  {journal} {\bibinfo  {journal} {Phys.
  Rev. A}\ }\textbf {\bibinfo {volume} {52}},\ \bibinfo {pages} {R2493}
  (\bibinfo {year} {1995})}\BibitemShut {NoStop}%
\bibitem [{\citenamefont {Nielsen}\ and\ \citenamefont
  {Chuang}(2000)}]{MikeandIke}%
  \BibitemOpen
  \bibfield  {author} {\bibinfo {author} {\bibfnamefont {M.~A.}\ \bibnamefont
  {Nielsen}}\ and\ \bibinfo {author} {\bibfnamefont {I.~L.}\ \bibnamefont
  {Chuang}},\ }\href@noop {} {\emph {\bibinfo {title} {Quantum {C}omputation
  and {Q}uantum {I}nformation}}}\ (\bibinfo  {publisher} {Cambridge Univ.
  Press},\ \bibinfo {address} {New York},\ \bibinfo {year} {2000})\BibitemShut
  {NoStop}%
\bibitem [{\citenamefont {{Cory et al.}}(1998)}]{CPMKLZHS}%
  \BibitemOpen
  \bibfield  {author} {\bibinfo {author} {\bibfnamefont {D.~G.}\ \bibnamefont
  {{Cory et al.}}},\ }\href@noop {} {\bibfield  {journal} {\bibinfo  {journal}
  {Phys. Rev. Lett.}\ }\textbf {\bibinfo {volume} {81}},\ \bibinfo {pages}
  {2152} (\bibinfo {year} {1998})}\BibitemShut {NoStop}%
\bibitem [{\citenamefont {Knill}\ \emph {et~al.}(2001)\citenamefont {Knill},
  \citenamefont {Laflamme}, \citenamefont {Martinez},\ and\ \citenamefont
  {Negrevergne}}]{KLMN}%
  \BibitemOpen
  \bibfield  {author} {\bibinfo {author} {\bibfnamefont {E.}~\bibnamefont
  {Knill}}, \bibinfo {author} {\bibfnamefont {R.}~\bibnamefont {Laflamme}},
  \bibinfo {author} {\bibfnamefont {R.}~\bibnamefont {Martinez}}, \ and\
  \bibinfo {author} {\bibfnamefont {C.}~\bibnamefont {Negrevergne}},\
  }\href@noop {} {\bibfield  {journal} {\bibinfo  {journal} {Phys. Rev. Lett.}\
  }\textbf {\bibinfo {volume} {86}},\ \bibinfo {pages} {5811} (\bibinfo {year}
  {2001})}\BibitemShut {NoStop}%
\bibitem [{\citenamefont {{Chiaverini et al.}}(2004)}]{Cetal}%
  \BibitemOpen
  \bibfield  {author} {\bibinfo {author} {\bibfnamefont {J.}~\bibnamefont
  {{Chiaverini et al.}}},\ }\href@noop {} {\bibfield  {journal} {\bibinfo
  {journal} {Nature}\ }\textbf {\bibinfo {volume} {432}},\ \bibinfo {pages}
  {602} (\bibinfo {year} {2004})}\BibitemShut {NoStop}%
\bibitem [{\citenamefont {Boulant}\ \emph {et~al.}(2005)\citenamefont
  {Boulant}, \citenamefont {Viola}, \citenamefont {Fortunato},\ and\
  \citenamefont {Cory}}]{BVFC}%
  \BibitemOpen
  \bibfield  {author} {\bibinfo {author} {\bibfnamefont {N.}~\bibnamefont
  {Boulant}}, \bibinfo {author} {\bibfnamefont {L.}~\bibnamefont {Viola}},
  \bibinfo {author} {\bibfnamefont {E.~M.}\ \bibnamefont {Fortunato}}, \ and\
  \bibinfo {author} {\bibfnamefont {D.~G.}\ \bibnamefont {Cory}},\ }\href@noop
  {} {\bibfield  {journal} {\bibinfo  {journal} {Phys. Rev. Lett.}\ }\textbf
  {\bibinfo {volume} {94}},\ \bibinfo {pages} {130501} (\bibinfo {year}
  {2005})}\BibitemShut {NoStop}%
\bibitem [{\citenamefont {{Schindler et al.}}(2011)}]{Schindleretal}%
  \BibitemOpen
  \bibfield  {author} {\bibinfo {author} {\bibfnamefont {P.}~\bibnamefont
  {{Schindler et al.}}},\ }\href@noop {} {\bibfield  {journal} {\bibinfo
  {journal} {Science}\ }\textbf {\bibinfo {volume} {332}},\ \bibinfo {pages}
  {1059} (\bibinfo {year} {2011})}\BibitemShut {NoStop}%
\bibitem [{\citenamefont {Moussa}\ \emph {et~al.}(2011)\citenamefont {Moussa},
  \citenamefont {Baugh}, \citenamefont {Ryan},\ and\ \citenamefont
  {Laflamme}}]{MBRL}%
  \BibitemOpen
  \bibfield  {author} {\bibinfo {author} {\bibfnamefont {O.}~\bibnamefont
  {Moussa}}, \bibinfo {author} {\bibfnamefont {J.}~\bibnamefont {Baugh}},
  \bibinfo {author} {\bibfnamefont {C.~A.}\ \bibnamefont {Ryan}}, \ and\
  \bibinfo {author} {\bibfnamefont {R.}~\bibnamefont {Laflamme}},\ }\href@noop
  {} {\bibfield  {journal} {\bibinfo  {journal} {Phys. Rev. Lett.}\ }\textbf
  {\bibinfo {volume} {107}},\ \bibinfo {pages} {160501} (\bibinfo {year}
  {2011})}\BibitemShut {NoStop}%
\bibitem [{\citenamefont {{Reed et al.}}(2012)}]{RDNSFGS}%
  \BibitemOpen
  \bibfield  {author} {\bibinfo {author} {\bibfnamefont {M.~D.}\ \bibnamefont
  {{Reed et al.}}},\ }\href@noop {} {\bibfield  {journal} {\bibinfo  {journal}
  {Nature}\ }\textbf {\bibinfo {volume} {482}},\ \bibinfo {pages} {382}
  (\bibinfo {year} {2012})}\BibitemShut {NoStop}%
\bibitem [{\citenamefont {{Yan et al.}}(2012)}]{TopoQEC}%
  \BibitemOpen
  \bibfield  {author} {\bibinfo {author} {\bibfnamefont {X.-C.}\ \bibnamefont
  {{Yan et al.}}},\ }\href@noop {} {\bibfield  {journal} {\bibinfo  {journal}
  {Nature}\ }\textbf {\bibinfo {volume} {482}},\ \bibinfo {pages} {489}
  (\bibinfo {year} {2012})}\BibitemShut {NoStop}%
\bibitem [{\citenamefont {Zhang}\ \emph {et~al.}(2012)\citenamefont {Zhang},
  \citenamefont {Laflamme},\ and\ \citenamefont {Suter}}]{ZLS}%
  \BibitemOpen
  \bibfield  {author} {\bibinfo {author} {\bibfnamefont {J.}~\bibnamefont
  {Zhang}}, \bibinfo {author} {\bibfnamefont {R.}~\bibnamefont {Laflamme}}, \
  and\ \bibinfo {author} {\bibfnamefont {D.}~\bibnamefont {Suter}},\
  }\href@noop {} {\bibfield  {journal} {\bibinfo  {journal} {Phys. Rev. Lett.}\
  }\textbf {\bibinfo {volume} {109}},\ \bibinfo {pages} {100503} (\bibinfo
  {year} {2012})}\BibitemShut {NoStop}%
\bibitem [{Note1()}]{Note1}%
  \BibitemOpen
  \bibinfo {note} {To avoid confusion with ``shared reference frames'' treated
  in \cite {BRS}, we use the term ``block'' as a synonym for ``frame''
  throughout this paper}\BibitemShut {NoStop}%
\bibitem [{\citenamefont {Laflamme}\ \emph {et~al.}(1996)\citenamefont
  {Laflamme}, \citenamefont {Miquel}, \citenamefont {Paz},\ and\ \citenamefont
  {Zurek}}]{LMPZ}%
  \BibitemOpen
  \bibfield  {author} {\bibinfo {author} {\bibfnamefont {R.}~\bibnamefont
  {Laflamme}}, \bibinfo {author} {\bibfnamefont {C.}~\bibnamefont {Miquel}},
  \bibinfo {author} {\bibfnamefont {J.~P.}\ \bibnamefont {Paz}}, \ and\
  \bibinfo {author} {\bibfnamefont {W.~H.}\ \bibnamefont {Zurek}},\ }\href@noop
  {} {\bibfield  {journal} {\bibinfo  {journal} {Phys. Rev. Lett.}\ }\textbf
  {\bibinfo {volume} {77}},\ \bibinfo {pages} {198} (\bibinfo {year}
  {1996})}\BibitemShut {NoStop}%
\bibitem [{\citenamefont {Sklar}(2001)}]{sklar}%
  \BibitemOpen
  \bibfield  {author} {\bibinfo {author} {\bibfnamefont {B.}~\bibnamefont
  {Sklar}},\ }\href@noop {} {\emph {\bibinfo {title} {Digital Communications:
  Fundamentals and Applications}}},\ \bibinfo {edition} {2nd}\ ed.\ (\bibinfo
  {publisher} {Prentice-Hall},\ \bibinfo {address} {Upper Saddle River, NJ},\
  \bibinfo {year} {2001})\BibitemShut {NoStop}%
\bibitem [{\citenamefont {Bregni}(2002)}]{bregni}%
  \BibitemOpen
  \bibfield  {author} {\bibinfo {author} {\bibfnamefont {S.}~\bibnamefont
  {Bregni}},\ }\href@noop {} {\emph {\bibinfo {title} {Synchronization of
  Digital Telecommunications Networks}}}\ (\bibinfo  {publisher} {John Wiley \&
  Sons},\ \bibinfo {address} {West Sussex, England},\ \bibinfo {year}
  {2002})\BibitemShut {NoStop}%
\bibitem [{Note2()}]{Note2}%
  \BibitemOpen
  \bibinfo {note} {One may argue that because of the no-cloning theorem,
  quantum communication would largely be restricted to point-to-point systems
  in which block synchronization is easier to establish than in broadcast
  networks. However, the very theorem forbids arguably the simplest and popular
  solution to any kind of error including synchronization errors, that is,
  requesting a perfect copy of lost information. Moreover, in quantum computing
  scenarios, because large-scale computing systems would benefit from internal
  communications between components and/or external communications, a
  fundamental building block of communications would become increasingly more
  relevant as the scale of realizable quantum computation grows.}\BibitemShut
  {Stop}%
\bibitem [{\citenamefont {Mercier}\ \emph {et~al.}(2010)\citenamefont
  {Mercier}, \citenamefont {Bhargava},\ and\ \citenamefont {Tarokh}}]{MBT}%
  \BibitemOpen
  \bibfield  {author} {\bibinfo {author} {\bibfnamefont {H.}~\bibnamefont
  {Mercier}}, \bibinfo {author} {\bibfnamefont {V.~K.}\ \bibnamefont
  {Bhargava}}, \ and\ \bibinfo {author} {\bibfnamefont {V.}~\bibnamefont
  {Tarokh}},\ }\href@noop {} {\bibfield  {journal} {\bibinfo  {journal} {{IEEE}
  Commun. Surveys Tutorials}\ }\textbf {\bibinfo {volume} {12}},\ \bibinfo
  {pages} {87} (\bibinfo {year} {2010})}\BibitemShut {NoStop}%
\bibitem [{\citenamefont {Huffman}\ and\ \citenamefont {Pless}(2003)}]{HP}%
  \BibitemOpen
  \bibfield  {author} {\bibinfo {author} {\bibfnamefont {W.~C.}\ \bibnamefont
  {Huffman}}\ and\ \bibinfo {author} {\bibfnamefont {V.}~\bibnamefont
  {Pless}},\ }\href@noop {} {\emph {\bibinfo {title} {Fundamentals of
  Error-Correcting Codes}}}\ (\bibinfo  {publisher} {Cambridge Univ. Press},\
  \bibinfo {address} {Cambridge},\ \bibinfo {year} {2003})\BibitemShut
  {NoStop}%
\bibitem [{\citenamefont {Calderbank}\ and\ \citenamefont {Shor}(1996)}]{CS}%
  \BibitemOpen
  \bibfield  {author} {\bibinfo {author} {\bibfnamefont {A.~R.}\ \bibnamefont
  {Calderbank}}\ and\ \bibinfo {author} {\bibfnamefont {P.~W.}\ \bibnamefont
  {Shor}},\ }\href@noop {} {\bibfield  {journal} {\bibinfo  {journal} {Phys.
  Rev. A}\ }\textbf {\bibinfo {volume} {54}},\ \bibinfo {pages} {1098}
  (\bibinfo {year} {1996})}\BibitemShut {NoStop}%
\bibitem [{\citenamefont {Steane}(1996)}]{Steane}%
  \BibitemOpen
  \bibfield  {author} {\bibinfo {author} {\bibfnamefont {A.~M.}\ \bibnamefont
  {Steane}},\ }\href@noop {} {\bibfield  {journal} {\bibinfo  {journal} {Phys.
  Rev. Lett.}\ }\textbf {\bibinfo {volume} {77}},\ \bibinfo {pages} {793}
  (\bibinfo {year} {1996})}\BibitemShut {NoStop}%
\bibitem [{\citenamefont {Grassl}\ and\ \citenamefont {Beth}(2000)}]{GB}%
  \BibitemOpen
  \bibfield  {author} {\bibinfo {author} {\bibfnamefont {M.}~\bibnamefont
  {Grassl}}\ and\ \bibinfo {author} {\bibfnamefont {T.}~\bibnamefont {Beth}},\
  }\href@noop {} {\bibfield  {journal} {\bibinfo  {journal} {Proc.\ R. Soc.\
  London Ser.\ A}\ }\textbf {\bibinfo {volume} {456}},\ \bibinfo {pages} {2689}
  (\bibinfo {year} {2000})}\BibitemShut {NoStop}%
\bibitem [{\citenamefont {Wilde}(2009)}]{Wilde}%
  \BibitemOpen
  \bibfield  {author} {\bibinfo {author} {\bibfnamefont {M.~M.}\ \bibnamefont
  {Wilde}},\ }\href@noop {} {\bibfield  {journal} {\bibinfo  {journal} {Phys.
  Rev. A}\ }\textbf {\bibinfo {volume} {79}},\ \bibinfo {pages} {062325}
  (\bibinfo {year} {2009})}\BibitemShut {NoStop}%
\bibitem [{\citenamefont {Steane}(1999)}]{Steane2}%
  \BibitemOpen
  \bibfield  {author} {\bibinfo {author} {\bibfnamefont {A.~M.}\ \bibnamefont
  {Steane}},\ }\href@noop {} {\bibfield  {journal} {\bibinfo  {journal} {{IEEE}
  Trans. Inf. Theory}\ }\textbf {\bibinfo {volume} {45}},\ \bibinfo {pages}
  {2492} (\bibinfo {year} {1999})}\BibitemShut {NoStop}%
\bibitem [{\citenamefont {Aly}\ \emph {et~al.}(2007)\citenamefont {Aly},
  \citenamefont {Klappenecker},\ and\ \citenamefont {Sarvepalli}}]{AKS}%
  \BibitemOpen
  \bibfield  {author} {\bibinfo {author} {\bibfnamefont {S.~A.}\ \bibnamefont
  {Aly}}, \bibinfo {author} {\bibfnamefont {A.}~\bibnamefont {Klappenecker}}, \
  and\ \bibinfo {author} {\bibfnamefont {P.~K.}\ \bibnamefont {Sarvepalli}},\
  }\href@noop {} {\bibfield  {journal} {\bibinfo  {journal} {{IEEE} Trans. Inf.
  Theory}\ }\textbf {\bibinfo {volume} {53}},\ \bibinfo {pages} {1183}
  (\bibinfo {year} {2007})}\BibitemShut {NoStop}%
\bibitem [{\citenamefont {Chang}\ \emph {et~al.}(1993)\citenamefont {Chang},
  \citenamefont {Sollenberger},\ and\ \citenamefont {Ariyavisitakul}}]{CSA}%
  \BibitemOpen
  \bibfield  {author} {\bibinfo {author} {\bibfnamefont {L.~F.}\ \bibnamefont
  {Chang}}, \bibinfo {author} {\bibfnamefont {N.~R.}\ \bibnamefont
  {Sollenberger}}, \ and\ \bibinfo {author} {\bibfnamefont {S.}~\bibnamefont
  {Ariyavisitakul}},\ }\href@noop {} {\bibfield  {journal} {\bibinfo  {journal}
  {{IEEE} Trans. Commun.}\ }\textbf {\bibinfo {volume} {41}},\ \bibinfo {pages}
  {22} (\bibinfo {year} {1993})}\BibitemShut {NoStop}%
\bibitem [{\citenamefont {Wu}\ and\ \citenamefont {Chang}(2010)}]{WC}%
  \BibitemOpen
  \bibfield  {author} {\bibinfo {author} {\bibfnamefont {Y.-W.}\ \bibnamefont
  {Wu}}\ and\ \bibinfo {author} {\bibfnamefont {S.-C.}\ \bibnamefont {Chang}},\
  }\href@noop {} {\bibfield  {journal} {\bibinfo  {journal} {{IEEE} Trans. Inf.
  Theory}\ }\textbf {\bibinfo {volume} {56}},\ \bibinfo {pages} {3786}
  (\bibinfo {year} {2010})}\BibitemShut {NoStop}%
\bibitem [{\citenamefont {Calderbank}\ \emph {et~al.}(1998)\citenamefont
  {Calderbank}, \citenamefont {Rains}, \citenamefont {Shor},\ and\
  \citenamefont {Sloane}}]{CRSSIEEE}%
  \BibitemOpen
  \bibfield  {author} {\bibinfo {author} {\bibfnamefont {A.~R.}\ \bibnamefont
  {Calderbank}}, \bibinfo {author} {\bibfnamefont {E.~M.}\ \bibnamefont
  {Rains}}, \bibinfo {author} {\bibfnamefont {P.~W.}\ \bibnamefont {Shor}}, \
  and\ \bibinfo {author} {\bibfnamefont {N.~J.~A.}\ \bibnamefont {Sloane}},\
  }\href@noop {} {\bibfield  {journal} {\bibinfo  {journal} {{IEEE} Trans. Inf.
  Theory}\ }\textbf {\bibinfo {volume} {44}},\ \bibinfo {pages} {1369}
  (\bibinfo {year} {1998})}\BibitemShut {NoStop}%
\bibitem [{\citenamefont {Grassl}\ \emph {et~al.}(1999)\citenamefont {Grassl},
  \citenamefont {Geiselmann},\ and\ \citenamefont {Beth}}]{GGB}%
  \BibitemOpen
  \bibfield  {author} {\bibinfo {author} {\bibfnamefont {M.}~\bibnamefont
  {Grassl}}, \bibinfo {author} {\bibfnamefont {W.}~\bibnamefont {Geiselmann}},
  \ and\ \bibinfo {author} {\bibfnamefont {T.}~\bibnamefont {Beth}},\ }in\
  \href@noop {} {\emph {\bibinfo {booktitle} {Lecture Notes in Computer
  Science}}},\ \bibinfo {series and number} {\bibinfo {series} {Proc. Applied
  Algebra, Algebraic Algorithm and Error-Correcting Codes}\ No.\ \bibinfo
  {number} {1719}},\ \bibinfo {editor} {edited by\ \bibinfo {editor}
  {\bibfnamefont {M.}~\bibnamefont {Fossorier}}, \bibinfo {editor}
  {\bibfnamefont {H.}~\bibnamefont {Imai}}, \bibinfo {editor} {\bibfnamefont
  {S.}~\bibnamefont {Lin}}, \ and\ \bibinfo {editor} {\bibfnamefont
  {A.}~\bibnamefont {Poli}}}\ (\bibinfo  {publisher} {Springer},\ \bibinfo
  {year} {1999})\ pp.\ \bibinfo {pages} {231--244}\BibitemShut {NoStop}%
\bibitem [{\citenamefont {Bartlett}\ \emph {et~al.}(2007)\citenamefont
  {Bartlett}, \citenamefont {Rudolph},\ and\ \citenamefont {Spekkens}}]{BRS}%
  \BibitemOpen
  \bibfield  {author} {\bibinfo {author} {\bibfnamefont {S.~D.}\ \bibnamefont
  {Bartlett}}, \bibinfo {author} {\bibfnamefont {T.}~\bibnamefont {Rudolph}}, \
  and\ \bibinfo {author} {\bibfnamefont {R.~W.}\ \bibnamefont {Spekkens}},\
  }\href@noop {} {\bibfield  {journal} {\bibinfo  {journal} {Rev. Mod. Phys.}\
  }\textbf {\bibinfo {volume} {79}},\ \bibinfo {pages} {555} (\bibinfo {year}
  {2007})}\BibitemShut {NoStop}%
\end{thebibliography}
\end{document}